# Computationally Efficient Molecular Integrals of Solid Harmonic Gaussian Orbitals Using Quantum Entanglement of Angular Momentum

*Hang Hu[1,2], Gilles Peslherbe[*1], Hsu Kiang Ooi[2] and Anguang Hu[*3]*

*[1]Centre for Research in Molecular Modeling and Department of Chemistry and Biochemistry, Concordia University, Montreal, Quebec, H4B 1R6, Canada*

*[2]Data Science for Complex Systems, National Research Council of Canada, 222 College Street Toronto, ON M5T 3J1*

*[3]Suffield Research Center, Defence Research Development Canada, PO Box 4000 Main Station, Medicine Hat, AB., T1A 8K6 Canada*

[*] Anguang.hu@drdc-rddc.gc.ca, Gilles.peslherbe@concordia.ca

**Abstract:**

Evaluating multi-center molecular integrals with Cartesian Gaussian-type basis sets has been a long-standing bottleneck in electronic structure theory calculation for solids and molecules. We have developed a vector-coupling and vector-uncoupling scheme to solve molecular Coulomb integrals with solid harmonics basis functions(SHGO). Solid harmonics are eigenstates of angular momentum, making it possible to factorize molecular integrals. By combining solid harmonic addition, differential and product rules, the computationally costly multi-center four-center integrals can be factored into an angular part and a radial component dependent on the atomic positions. The potential speed-up ratio in evaluating molecular nuclear Coulomb integrals in our method can reach up to four orders of magnitude for atomic orbitals with high angular momentum quantum numbers. The foundation underpinning the mathematical efficiency is the quantum angular momentum theory, where both vector-coupling and vector-uncoupling schemes correspond to unitary Clebsch-Gordan transformations that act on quantum angular momentum states, influencing their degree of entanglement. By incorporating quantum angular momentum through these transformations, the entanglement of the states can be reduced, and the less entanglement there is for a quantum system, the easier it is to simulate. The highly efficient method unveiled here opens new avenues for accelerated material and molecule design and discovery.

In the quantum mechanic description of atoms, orbitals, with well-defined energy levels, angular momentum and magnetic momentum, are the building blocks of the electronic structure [1]. This conceptual framework laid the groundwork for contemporary computational chemistry methods such as Hartree-Fock and density-functional theory to characterize orbitals through mathematical wave functions. The wave functions are typically a product of a radial component, which defines the energy and an eigenfunction of the angular momentum operator for the angular part of the atomic orbitals [2-4]. For systems with rotational symmetry, angular momentum plays a central role in simplifying complicated mathematical modelling into smaller factors[5, 6]. In general, these simple factors may be divided into two types. The first type is invariant under rotation, mainly determined by the precise physical characteristics of the quantum system under consideration. In contrast, the second type depends solely on the system's rotational properties. Therefore, the second type of factor can be precisely expressed as a function of angular momenta, laying the foundation for a general theory of angular momentum algebra. The elegant computational methods derived from this theory were widely applied to many problems, such as atomic, molecular, and nuclear spectroscopies, as well as nuclear reactions[7, 8]. For computational quantum chemistry, the most popular method still relies on Cartesian Gaussian orbitals (CGOs). Unlike the CGOs, solid harmonic Gaussian orbitals (SHGO) can naturally describe the spatial dependence of wavefunctions for systems with spherical symmetry and separate the radial component from the angular part. There were some past developments for one-electron molecular integrals using SHGO[9-11]. However, to the best of our knowledge, efficient calculations of molecular Coulomb integrals with SHGOs have not yet been reported [12-14]. In this paper, we present an approach for the efficient evaluation of molecular integrals with SHGOs using a combination of vector-coupling and vector-uncoupling schemes of angular momenta. We demonstrate that the highly efficient calculation of molecular Coulomb integrals with SHGOs arises from simpler mathematical expressions and their quantum nature (the entanglement degree of orbital angular momentum states).

Taking the overlap and nuclear Coulomb attraction integrals as paradigms in this paper, we show a detailed derivation of an efficient to evaluate expressions for molecular integrals using SHGOs and relevant vector-coupling and vector-uncoupling schemes of angular momenta. In general, SHGOs are defined as

$$\chi_{m_a}^{l_a\vec{a}\alpha}\left(\vec{r}\right)=N\left(l_a,\alpha\right)\mathcal{Y}_{m_a}^{l_a}\left(\vec{r}-\vec{a}\right)e^{-\alpha\left|\vec{r}-\vec{a}\right|^2} \quad (1),$$

where $\vec{a}$ is the orbital atomic center, $\alpha$ the Gaussian exponent, $N\left(l_a,\alpha\right)$ the normalization constant, and $\mathcal{Y}_{m_a}^{l_a}\left(\vec{r}-\vec{a}\right)$ are solid harmonics (using the same notation from reference 6), a solution of the Laplace equation. Solid harmonics are related to the spherical harmonics as [6]

$$\mathcal{Y}_m^l\left(\vec{r}\right)=\sqrt{\frac{4\pi}{2l+1}}r^lY_{lm}\left(\theta,\phi\right) \quad (2),$$

where $Y_{lm}\left(\theta,\phi\right)$ are the spherical harmonics with the phase convention of Condon and Shortley [8], $l$ and $m$ are the orbital angular momentum and magnetic quantum numbers, respectively. With the Hobson theorem of solid harmonics[15] SHGOs can be further simplified using solid harmonic derivatives with respect to the orbital atomic center,

$$\chi_{m_a}^{l_a\vec{a}\alpha}\left(\vec{r}\right)=N\left(l_a,\alpha\right)\left(\frac{1}{2\alpha}\right)^l\mathcal{Y}_{m_a}^{l_a}\left(\nabla_{\vec{a}}\right)e^{-\alpha\left|\vec{r}-\vec{a}\right|^2} \quad (3).$$

Without the normalization constants, the general expression for a molecular overlap integral is of the form

$$\boldsymbol{S}_{\alpha\beta}^{\vec{a}\vec{b}}\left(\boldsymbol{l}_a\boldsymbol{m}_a,\boldsymbol{l}_b\boldsymbol{m}_b\right)=\left(\frac{1}{2\alpha}\right)^{l_a}\left(\frac{1}{2\beta}\right)^{l_b}\mathcal{Y}_{m_a}^{l_a}\left(\nabla_{\vec{a}}\right)\mathcal{Y}_{m_b}^{l_b}\left(\nabla_{\vec{b}}\right)\int \boldsymbol{e}^{-\alpha\left|\vec{r}-\vec{a}\right|^2}\boldsymbol{e}^{-\beta\left|\vec{r}-\vec{b}\right|^2}\boldsymbol{d}\tau \quad (4).$$

Using the Gaussian product rule,

$$\boldsymbol{e}^{-\alpha\left|\vec{r}-\vec{a}\right|^2}\boldsymbol{e}^{-\beta\left|\vec{r}-\vec{b}\right|^2}=\boldsymbol{e}^{-\frac{\alpha\beta}{\alpha+\beta}\left|\vec{a}-\vec{b}\right|^2}\boldsymbol{e}^{-(\alpha+\beta)\left|\vec{r}-\vec{P}\right|^2} \quad (5),$$

where $\vec{P}=\dfrac{\alpha\vec{a}+\beta\vec{b}}{\alpha+\beta}$, and integration yields overlap integrals given by

$$\boldsymbol{S}_{\alpha\beta}^{\vec{a}\vec{b}}\left(\boldsymbol{l}_a\boldsymbol{m}_a,\boldsymbol{l}_b\boldsymbol{m}_b\right)=\left(\frac{1}{2\alpha}\right)^{l_a}\left(\frac{1}{2\beta}\right)^{l_b}\left(\frac{\pi}{\alpha+\beta}\right)^{\frac{3}{2}}\mathcal{Y}_{m_a}^{l_a}\left(\nabla_{\vec{a}}\right)\mathcal{Y}_{m_b}^{l_b}\left(\nabla_{\vec{b}}\right)\boldsymbol{e}^{-\frac{\alpha\beta}{\alpha+\beta}\left|\vec{a}-\vec{b}\right|^2} \quad (6).$$

The overlap integral in Eq. (6) can be evaluated by applying the differential and product rules of solid harmonic derivative[9]. This approach is similar to the previously derived vector-uncoupling scheme of angular momenta because there are no terms related to the $\boldsymbol{l}_a+\boldsymbol{l}_b$ quantum number. However, this is suboptimal in terms of computational efficiency, with too many harmonic derivatives acting on different atomic centers[9], resulting in deep angular momentum entanglement of the atomic orbitals. Our new approach can eliminate the harmonics derivatives with the addition of angular momentum [6]. With this addition, we shift orbital atomic centers $\vec{a}$ and $\vec{b}$ to the same center $\vec{P}$ of Gaussian $\boldsymbol{e}^{-(\alpha+\beta)\left|\vec{r}-\vec{P}\right|^2}$, and obtain the following equation[16]

$$\mathcal{Y}_{m_a}^{l_a}(\vec{r}-\vec{a})=\sum_{l_1=0}^{l_a}\left(\frac{\beta}{\alpha+\beta}\right)^{l_a-l_1}\sum_{m_1=-l_1}^{l_1}\varepsilon_{l_a l_1 m_a m_1}\mathcal{Y}_{m_a-m_1}^{l_a-l_1}\left(\vec{a}-\vec{b}\right)\mathcal{Y}_{m_1}^{l_1}\left(\vec{r}-\vec{P}\right) \quad (7).$$

Here, the vector-coupling coefficients of angular momenta are given by [16]

$$\varepsilon_{l_1 l_2 m_1 m_2} = \sqrt{\binom{l_1+m_1}{l_2+m_2}\binom{l_1-m_1}{l_2-m_2}} \quad (8),$$

and are also Clebsch-Gordan (CG) transformation coefficients[9]. Therefore, a two-center overlap integral is transformed into a single-center integral,

$$\begin{aligned} S_{\alpha\beta}^{\vec{a}\vec{b}}\left(l_a m_a, l_b m_b\right) = e^{-\frac{\alpha\beta}{\alpha+\beta}\left|\vec{a}-\vec{b}\right|^2} \sum_{l_1=0}^{l_a}\left(\frac{\beta}{\alpha+\beta}\right)^{l_a-l_1} \sum_{m_1=-l_1}^{l_1} \varepsilon_{l_a l_1 m_a m_1} \mathcal{Y}_{m_a-m_1}^{l_a-l_1}\left(\vec{a}-\vec{b}\right) \\ \sum_{l_2=0}^{l_b}\left(\frac{\alpha}{\alpha+\beta}\right)^{l_b-l_2} \sum_{m_2=-l_2}^{l_2} \varepsilon_{l_b l_2 m_b m_2} \mathcal{Y}_{m_b-m_2}^{l_b-l_2}\left(\vec{b}-\vec{a}\right) \\ \int \mathcal{Y}_{m_1}^{l_1}\left(\vec{r}-\vec{P}\right) \mathcal{Y}_{m_2}^{l_2}\left(\vec{r}-\vec{P}\right) e^{-(\alpha+\beta)\left|\vec{r}-\vec{P}\right|^2} d\tau \end{aligned} \quad (9).$$

With the orthonormality of solid harmonics, we obtain

$$\begin{aligned} S_{\alpha\beta}^{\vec{a}\vec{b}}\left(l_a m_a; l_b m_b\right) = 4\pi\left(\frac{\pi}{2}\right)^{\frac{1}{2}} e^{-\frac{\alpha\beta}{\alpha+\beta}\left|\vec{a}-\vec{b}\right|^2} \sum_{l_1=0}^{\min(l_a,l_b)}\left(\frac{\beta}{\alpha+\beta}\right)^{l_a-l_1}\left(\frac{\alpha}{\alpha+\beta}\right)^{l_b-l_1}\left(\frac{1}{2l_1+1}\right)(-1)^{l_b-l_1} \\ \frac{(2l_1+1)!!}{\left[2\left(\alpha+\beta\right)\right]^{\frac{2l_1+3}{2}}} \sum_{m_1=-l_1}^{l_1} \varepsilon_{l_a l_1 m_a m_1} \varepsilon_{l_b l_1 m_b m_1} \mathcal{Y}_{m_a-m_1}^{l_a-l_1}\left(\vec{a}-\vec{b}\right) \mathcal{Y}_{m_b-m_1}^{l_b-l_1}\left(\vec{a}-\vec{b}\right) \end{aligned} \quad (10).$$

The summation in the second line of Eq. (10) is independent of orbitals and determined only by the rotational properties of angular momenta for an overlap integral. The last line in Eq. (9) is an elegant solution to demonstrate that quantum angular momenta of the same atomic center can not entangle with each other due to their orthonormality. It indicates that if a quantum system does not have much entanglement, it will be much easier to simulate. As a result, in the same way, the nuclear Coulomb attraction integrals can be directly calculated by simply introducing a nuclear Coulomb operator located at the center $\vec{c}$, into the integral of the last line in Eq. (9),

$$
\boldsymbol{N}_{\alpha\beta}^{\vec{a}\vec{b}\vec{c}}\left(\boldsymbol{l}_a\boldsymbol{m}_a,\boldsymbol{l}_b\boldsymbol{m}_b\right)=\boldsymbol{e}^{-\frac{\alpha\beta}{\alpha+\beta}\left|\vec{a}-\vec{b}\right|^2}\sum_{l_1=0}^{l_a}\left(\frac{\beta}{\alpha+\beta}\right)^{l_a-l_1}\sum_{m_1=-l_1}^{l_1}\varepsilon_{l_a l_1 m_a m_1}\mathcal{Y}_{m_a-m_1}^{l_a-l_1}\left(\vec{a}-\vec{b}\right)
$$

$$
\sum_{l_2=0}^{l_b}\left(\frac{\alpha}{\alpha+\beta}\right)^{l_b-l_2}\sum_{m_2=-l_2}^{l_2}\varepsilon_{l_b l_2 m_b m_2}\mathcal{Y}_{m_b-m_2}^{l_b-l_2}\left(\vec{b}-\vec{a}\right) \qquad (11).
$$

$$
\int\frac{\mathcal{Y}_{m_1}^{l_1}\left(\vec{r}-\vec{P}\right)\mathcal{Y}_{m_2}^{l_2}\left(\vec{r}-\vec{P}\right)\boldsymbol{e}^{-(\alpha+\beta)\left|\vec{r}-\vec{P}\right|^2}}{\left|\vec{r}-\vec{c}\right|}\boldsymbol{d}\tau
$$

The last line of Eq. (11) clearly expresses the nuclear Coulomb attraction integral as the interaction between Gaussian distributed angular momenta at the center $\vec{P}$ and the nuclear Coulomb potential at the center $\vec{c}$ . The well-known formula of two solid harmonic products may be applied to simplify the calculation,

$$
\mathcal{Y}_{m_1}^{l_1}\left(\vec{r}-\vec{P}\right)\mathcal{Y}_{m_2}^{l_2}\left(\vec{r}-\vec{P}\right)=
$$

$$
\sum_{|l_1-l_2|}^{l_1+l_2}\sum_{m_p=-l_p}^{l_p}\left[\frac{(2l_1+1)(2l_1+1)}{4\pi\left(2l_p+1\right)}\right]\begin{pmatrix} l_1 & l_1 & l_p \\ 0 & 0 & 0 \end{pmatrix}\begin{pmatrix} l_1 & l_2 & l_p \\ m_1 & m_2 & m_p \end{pmatrix}\left|\vec{r}-\vec{P}\right|^{l_1+l_2-l_p}\mathcal{Y}_{m_p}^{l_p}\left(\vec{r}-\vec{P}\right) \qquad (12),
$$

where $\begin{pmatrix} l_1 & l_2 & l_p \\ m_1 & m_2 & m_p \end{pmatrix}$ is the Wigner 3-j symbol. Note that the only $l_p$ values on the right should satisfy $l_1+l_2+l_p=$ even integers. Combining CG coefficients $\varepsilon_{l_a l_1 m_a m_1}$ , $\varepsilon_{l_b l_2 m_b m_2}$ , $\mathcal{Y}_{m_a-m_1}^{l_a-l_1}\left(\vec{a}-\vec{b}\right)$ and $\mathcal{Y}_{m_b-m_2}^{l_b-l_2}\left(\vec{b}-\vec{a}\right)$ in Equation (11) with the Wigner 3-j symbols, we have a three-dimensional array $\left(l_1+l_2,l_p,m_p\right)_{\vec{b}l_b m_b}^{\vec{a}l_a m_a}$ , which is independent of orbitals. Thus, the nuclear Coulomb integral can be further simplified to

$$
\begin{aligned}
&\int \frac{\left|\boldsymbol{r}-\boldsymbol{P}\right|^{2\left[\frac{l_1+l_2-l_p}{2}\right]} \mathcal{Y}_{m_p}^{l_p}\left(\vec{\boldsymbol{r}}-\vec{\boldsymbol{P}}\right) e^{-(\alpha+\beta)\left|\vec{\boldsymbol{r}}-\vec{\boldsymbol{P}}\right|^2}}{\left|\vec{\boldsymbol{r}}-\vec{\boldsymbol{c}}\right|} d\tau \\
&= \left(-\frac{d}{d(\alpha+\beta)}\right)^{\left[\frac{l_1+l_2-l_p}{2}\right]} \int \frac{\mathcal{Y}_{m_p}^{l_p}\left(\vec{\boldsymbol{r}}-\vec{\boldsymbol{P}}\right) e^{-(\alpha+\beta)\left|\vec{\boldsymbol{r}}-\vec{\boldsymbol{P}}\right|^2}}{\left|\vec{\boldsymbol{r}}-\vec{\boldsymbol{c}}\right|} d\tau \\
&= 2\pi \mathcal{Y}_{m_p}^{l_p}\left(\vec{\boldsymbol{c}}-\vec{\boldsymbol{P}}\right)\left(-\frac{d}{d(\alpha+\beta)}\right)^{\left[\frac{l_1+l_2-l_p}{2}\right]} \int_0^1 du\, u^{2l_p} \frac{e^{-(\alpha+\beta)u^2\left|\vec{\boldsymbol{c}}-\vec{\boldsymbol{P}}\right|^2}}{\alpha+\beta}
\end{aligned} \qquad (13),
$$

where $\left[\frac{l_1+l_2-l_p}{2}\right]$ is a positive integer. Compared with all available integrals of the same type with either SHGOs or CGOs, Eq. (13) is the simplest expression for general nuclear Coulomb attraction integrals. The final expression combines Eqs. (11), (12) and (13) and contains three parts. The first one is a three-dimensional array $\left(l_1+l_2, l_p, m_p\right)_{\vec{b}l_b m_b}^{\vec{a}l_a m_a}$ and is independent of Gaussian exponents. By separating out the Gaussian exponents from this part, we can calculate it initially and reuse it during the later steps, and it can even be applied to the two-electron Coulomb integral. The second part involves $\mathcal{Y}_{m_p}^{l_p}\left(\vec{\boldsymbol{c}}-\vec{\boldsymbol{P}}\right)$, which is dependent on Gaussian orbitals. The third part is related to the Boys function, $\left(-\frac{d}{d(\alpha+\beta)}\right)^{\left[\frac{l_1+l_2-l_p}{2}\right]} \int_0^1 du\, u^{2l_p} \frac{e^{-(\alpha+\beta)u^2\left|\vec{\boldsymbol{c}}-\vec{\boldsymbol{P}}\right|^2}}{\alpha+\beta}$ dependent on Gaussian exponents [17], which can be simplified as $\left(\frac{d}{dx}\right)^n \frac{e^{-x}}{x}$. Table 1 shows a comparison of the computational cost of evaluating the nuclear Coulomb attraction integral with CGOs and SHGOs [18, 19]. The dominant computational cost with CGOs scales as $L^7P^2$ ($L$ is the highest angular momentum number, and $P$ is the number of primitive Gaussians). Previous SHGO methods with only the vector-uncoupling scheme of angular momentum, as implemented in the ParaGauss package, have a similar computational cost of $L^6P^2$ [14]. This relatively high cost is the result of three-fold summations over angular momentum $l$ and magnetic quantum $m$, respectively[20]. The interaction of two orbital angular momenta from different centers with a nuclear Coulomb potential can generate a lot of quantum entanglement under the vector-uncoupling scheme. As a result, the vector-uncoupling scheme of angular momenta is not suitable for multi-center molecular integrals. This can be the main reason why two-electron Coulomb integrals with SHGOs have not been implemented in earlier quantum chemistry packages. Figure 1 shows the potential computational overall speed-up ratio achieved using our approach with SHGOs. For calculations of CGOs using $L$=7 ($h$ orbital) as the highest angular momentum number and $P$=10 for the number of primitive Gaussians, the speed-up ratio is roughly about 14,600 times, based on an estimate from Table 1. This represents a computational efficiency gain of 4 orders of magnitude.

Table 1. The cost associated with computing nuclear attraction integrals

| Calculation steps with CGOs[19] | Cost | Calculation steps with SHGOs[21] | Cost | This work with SHGOs | Cost |
|---|---|---|---|---|---|
| Boys functions | $LP^2$ | Solid Harmonics | $L^2$ | Solid harmonics | $L^2P^2$ |
| Hermite integrals | $L^4P^2$ | Solid Harmonics | $L^2P^2$ | Boys functions | $LP^2$ |
| Expansion coefficients | $L^2P^2$ | Differential rule | $L^4P^2$ | | |
| Cartesian integrals | $L^7P^2$ | Product rule | $L^6P^2$ | | |
| Primitive Contraction | $L^3P^2$ | Boys functions | $LP^2$ | | |
| Solid harmonics | $L^5$ | | | | |

$L$ is the highest orbital angular momentum number of the Gaussian orbitals and
$P$ is the number of primitive Gaussian functions.

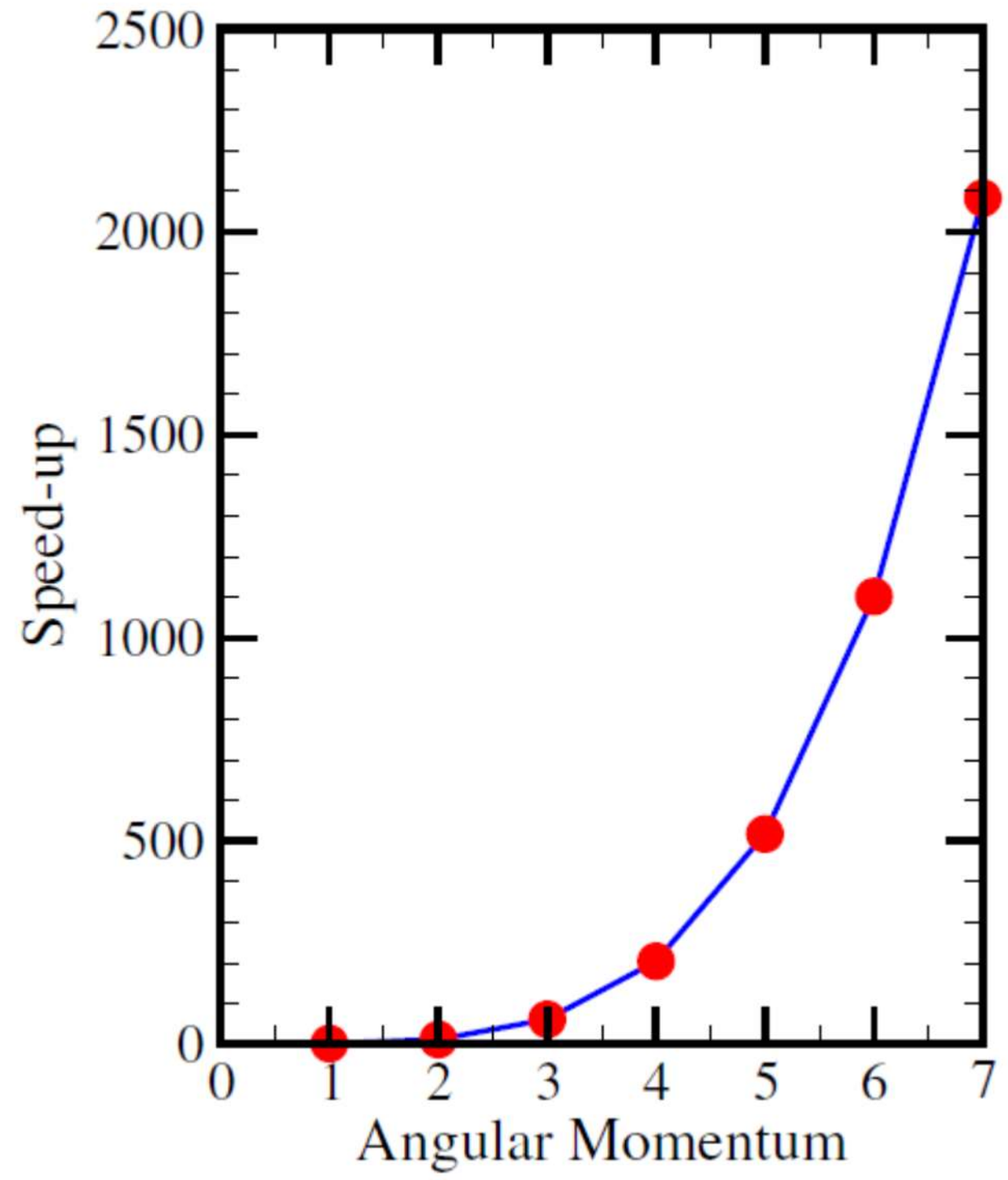

Figure 1. Computational speed-up ratio in the calculation of nuclear attraction Coulomb integrals as a function of highest orbital angular momentum number.

For a general two-electron Coulomb integral, there are four Gaussian exponents $\alpha$, $\beta$, $\delta$, and $\lambda$ located at four atomic orbital centers $\vec{\boldsymbol{a}}$, $\vec{\boldsymbol{b}}$, $\vec{\boldsymbol{c}}$, and $\vec{\boldsymbol{d}}$ respectively. We can transform this two-electron, four-center Coulomb integral into a two-center one with the centres located at $\vec{\boldsymbol{P}} = \frac{\alpha\vec{\boldsymbol{a}} + \beta\vec{\boldsymbol{b}}}{\alpha+\beta}$ and $\vec{\boldsymbol{Q}} = \frac{\delta\vec{\boldsymbol{c}} + \lambda\vec{\boldsymbol{d}}}{\delta+\lambda}$ in terms of four Gaussian distributed angular momenta based on Eq. (7). It leads to the simple expression as follows,

$$I\left(\vec{r}_1-\vec{P},\vec{r}_2-\vec{Q}\right)_{\varsigma l_3 m_3 l_4 m_4}^{\xi l_1 m_1 l_2 m_2} =$$

$$\int\int \frac{\mathcal{Y}_{m_1}^{l_1}\left(\vec{r}_1-\vec{P}\right)\mathcal{Y}_{m_2}^{l_2}\left(\vec{r}_1-\vec{P}\right)\mathcal{Y}_{m_3}^{l_3}\left(\vec{r}_2-\vec{Q}\right)\mathcal{Y}_{m_4}^{l_4}\left(\vec{r}_2-\vec{Q}\right)e^{-\xi\left|\vec{r}_1-\vec{P}\right|^2}e^{-\varsigma\left|\vec{r}_2-\vec{Q}\right|^2}}{\left|\vec{r}_1-\vec{r}_2\right|}d\tau_1 d\tau_2 \qquad (14),$$

where $\xi=\alpha+\beta$ and $\varsigma=\delta+\gamma$. Applying the same transformation used in Equation (12), we can rewrite equation 14 as:

$$I\left(\vec{r}_1-\vec{P},\vec{r}_2-\vec{Q}\right)_{\varsigma l_q m_q}^{\xi l_p m_p} =$$

$$\int\int \frac{\left|\vec{r}_1-\vec{P}\right|^{2\left[\frac{l_1+l_2-l_p}{2}\right]}\mathcal{Y}_{m_p}^{l_p}\left(\vec{r}_1-\vec{P}\right)\left|\vec{r}_2-\vec{Q}\right|^{2\left[\frac{l_3+l_4-l_q}{2}\right]}\mathcal{Y}_{m_q}^{l_q}\left(\vec{r}_2-\vec{Q}\right)e^{-\xi\left|\vec{r}_1-\vec{P}\right|^2}e^{-\varsigma\left|\vec{r}_2-\vec{Q}\right|^2}}{\left|\vec{r}_1-\vec{r}_2\right|}d\tau_1 d\tau_2 \qquad (15)$$

$$=\left(-\frac{d}{d\xi}\right)^{2\left[\frac{l_1+l_2-l_p}{2}\right]}\left(-\frac{d}{d\varsigma}\right)^{2\left[\frac{l_3+l_4-l_q}{2}\right]}\int\int \frac{\mathcal{Y}_{m_p}^{l_p}\left(\vec{r}_1-\vec{P}\right)\mathcal{Y}_{m_q}^{l_q}\left(\vec{r}_2-\vec{Q}\right)e^{-\xi\left|\vec{r}_1-\vec{P}\right|^2}e^{-\varsigma\left|\vec{r}_2-\vec{Q}\right|^2}}{\left|\vec{r}_1-\vec{r}_2\right|}d\tau_1 d\tau_2$$

In this simplified process, there are two three-dimensional arrays $\left(l_1+l_2,l_p,m_p\right)_{\vec{b}l_b m_b}^{\vec{a}l_a m_a}$ and $\left(l_3+l_4,l_q,m_q\right)_{\vec{d}l_d m_d}^{\vec{c}l_c m_c}$, which are independent of Gaussian orbitals and share all relevant integrals including nuclear Coulomb ones. As a result, we have the simple expression of $I\left(\vec{r}_1-\vec{P},\vec{r}_2-\vec{Q}\right)_{\varsigma l_q m_q}^{\xi l_p m_p}$,

$$
\begin{aligned}
& I\left(\vec{r}_1-\vec{P},\vec{r}_2-\vec{Q}\right)_{\varsigma l_q m_q}^{\xi l_p m_p}=2\pi^{\frac{5}{2}}\left(-\frac{d}{d\xi}\right)^{2\left[\frac{l_1+l_2-l_p}{2}\right]}\left(-\frac{d}{d\varsigma}\right)^{2\left[\frac{l_3+l_4-l_q}{2}\right]}\frac{1}{\xi\varsigma\sqrt{\xi+\varsigma}}\left(\frac{\varsigma}{\xi+\varsigma}\right)^{l_p}\left(\frac{\xi}{\xi+\varsigma}\right)^{l_q} \\
& \sum_{l_t=0}^{\min(l_p,l_q)}\left(-\frac{\xi+\varsigma}{2\xi\varsigma}\right)^{l_t}(2l_t-1)!!\int_0^1 u^{2\left(l_p+l_q-l_t\right)}e^{-\frac{\xi\varsigma}{\xi+\varsigma}\left|\vec{P}-\vec{Q}\right|^2u^2}du \\
& \sum_{m_t=-l_t}^{l_t}\varepsilon_{l_p l_t m_p m_t}\varepsilon_{l_q l_t m_q m_t}\mathcal{Y}_{m_q-m_t}^{l_q-l_t}\left(\vec{P}-\vec{Q}\right)\mathcal{Y}_{m_p-m_t}^{l_p-l_t}\left(\vec{Q}-\vec{P}\right)
\end{aligned} \tag{16}
$$

Equation 16 involves two 5-dimensional array factors. The first array $\left(l_1+l_2,l_p,l_3+l_4,l_q,l_t\right)_{\vec{Q}\varsigma}^{\vec{P}\xi}$ corresponds to the first two lines in Equation (16). The second array $\left(l_p,m_p,l_q,m_q,l_t\right)_{\vec{Q}}^{\vec{P}}$ is due to the summation over $m_t$ of the last line in Equation (16). The final expression of the general two-electron Coulomb integral is given by the summation of four array factors, $\left(l_1+l_2,l_p,m_p\right)_{\vec{b}l_b m_b}^{\vec{a}l_a m_a}$, $\left(l_3+l_4,l_q,m_q\right)_{\vec{d}l_d m_d}^{\vec{c}l_c m_c}$, $\left(l_p,m_p,l_q,m_q,l_t\right)_{\vec{Q}}^{\vec{P}}$, and $\left(l_1+l_2,l_p,l_3+l_4,l_q,l_t\right)_{\varsigma\vec{Q}}^{\xi\vec{P}}$, with a scaling of $L^7P^4$ where $L$ is the highest angular momentum and $P$ is the number of the primitive Gaussian functions. The actual scaling may be much smaller when we apply the constraints where $l_1+l_2+l_p$ and $l_3+l_4+l_q$ must be even positive integers. The constraints also applies to the summation of the second line in Equation (16), together with $\left|l_1-l_2\right|\le l_p\le l_1+l_2$ and $\left|l_3-l_4\right|\le l_q\le l_3+l_4$. Moreover, symmetry conditions for the Wigner 3-j symbol also reduce the number of nonzero terms in the array factors $\left(l_1+l_2,l_p,m_p\right)_{\vec{b}l_b m_b}^{\vec{a}l_a m_a}$ and $\left(l_3+l_4,l_q,m_q\right)_{\vec{d}l_d m_d}^{\vec{c}l_c m_c}$, resulting in significant computational savings. More importantly, calculations are all exact without any approximation. Compared with the dominant computational cost such as $L^{13}P^4$, $L^9P^4$, and $L^6P^4$ associated with two-electron Coulomb integrals with CGOs[19], the speed-up ratio may increase by several orders of magnitude.

In conclusion, this paper presents the development of a framework to formulate and efficiently evaluate molecular integrals in Gaussian orbitals using the entanglement of angular momentum in quantum mechanics. The addition of solid harmonics and the Clebsch-Gordan transformation allows multi-center molecular integrals to be converted into one- or two-center integrals, independent of the original angular momenta of the Gaussian orbitals. Generally, the less entanglement in a quantum system, the easier it is to simulate. The Clebsch-Gordan transformation for the addition of solid harmonics can significantly reduce the entanglement degree of angular momentum of molecular integrals. Consequently, molecular integrals can be efficiently calculated with a computational speed-up of several orders of magnitude, depending on the highest orbital angular momentum quantum numbers. The formalism for efficient evaluation of molecular integrals with solid harmonic Gaussian orbitals developed in this paper could thus pave the way for quantum computational chemistry investigations of realistic systems on reasonable timescales. Finally, the computational approach developed in this paper may also be applied for simulations of complex forms of quantum entanglement, such as those arising from the interaction between orbital angular momenta of light described as Laguerre-Gaussian modes and quantum states of matter for the development of the photonic quantum computer[22]. As part of the future work, we plan to implement the developed calculation scheme into ParaGauss, a freely available package under the GNU General Public Licence, in which SHGOs are already used..